\documentclass[aps,preprint,showpacs]{revtex4}
\usepackage{latexsym}
\usepackage{graphicx}

\begin{document}
\preprint{\vbox{\hbox{\tt gr-qc/0612023}}}

\title{Future singularity free accelerating expansion with the
  modified Poisson brackets} 
\author{Wontae Kim}\email{wtkim@sogang.ac.kr}
\affiliation{Department of Physics, Sogang University, C.P.O. Box 1142, Seoul 100-611, Korea}
\author{Edwin J. Son}\email{ejson@yonsei.ac.kr}
\affiliation{Institute of Physics and Applied Physics
 and Natural Science Research Institute,
 Yonsei University, Seoul 120-749, Korea}

\date{\today}

\begin{abstract}
We show that the second accelerating expansion of the universe appears
smoothly from the decelerating phase, which follows the initial inflation,
in the two-dimensional soluble semi-classical dilaton gravity 
along with the modified Poisson
brackets with noncommutativity between the relevant fields. 
This is contrast to the fact that the ordinary solution of
the equations of motion following from the conventional 
Poisson algebra describes permanent accelerating
universe without any phase change. In this modified model, 
it turns out that the noncommutative Poisson algebra is responsible
for the remarkable phase transition to the second accelerating expansion.
\end{abstract}

\maketitle

\section{Introduction}
\label{sec:intro}
It has been proposed that the recent intriguing accelerating expansion of the
universe~\cite{perlmutter} is of relevance to the curious
behavior of dark energy described by the negative
equation-of-state parameter~\cite{caldwell}. In the
well-known Friedmann equation, it is easily expected for our 
universe to show the decelerating expansion if 
the positive energy condition has been met
as seen from $3\ddot{a}/a=-4 \pi G (\rho + 3 p)$, where $a$ is the scale
factor, and $\rho$ and $p$ are the energy density and the pressure,
respectively. So, the dark energy, which is responsible for the
accelerating expansion of the universe, is defined by the state
parameter, $\omega (\equiv p/\rho) < -1/3$ and even more $\omega < -1$ for
the phantom energy to compensate ordinary matters. Note that the
density of the dark energy is assumed to be positive so that the
pressure should be negative, which naturally results in the negative
state parameter~\cite{carroll}. 
 
On the other hand, cosmological problems are usually hard to solve
exactly and thus often considered in some simplified models in order
to get some clues and insights for solving them. One such model is
the exactly soluble two-dimensional
gravity~\cite{cghs,str,rst,mrcm,bpp,rt,kv,bc,bk:1}, which has
attracted much attention in the study of cosmology in various
aspects~\cite{gv,rey,bk,ky:dg,ky:bd,cdkz} (for a review
of two or higher dimensional dilaton gravity, see Ref.~\cite{no}).
From this perspective, it would be interesting to study whether the
model can show accelerating expansion of the universe after the
decelerating expansion or not. A recent work~\cite{ky:ncdc} shows that
it may be possible to obtain the transition from the decelerating
universe to the accelerating universe by assuming noncommutativity
during the finite time. However, this model essentially encounters the
future singularity in a finite proper time unless an appropriate regular
geometry is patched. So, we would like to obtain cosmological
solutions to describe the future-singularity-free accelerating
universe without patching after the initial acceleration corresponding
to the first inflation, whose whole profile is essentially similar to
our universe chronologically. For this purpose, we add two local
counter terms with the Polyakov action of the conformal anomaly in the
semi-classical action, and impose some modified Poisson brackets with
noncommutativity between the relevant fields. This process naturally
yields the modified set of semi-classical equations of motion
involving the noncommutative parameter, and remarkably leads us to
have the desired solutions.

In the next section, we shall introduce the
Callan-Giddings-Harvey-Strominger model~\cite{cghs,rst} coupled to
massless conformal matter fields and its quantum correction expressed
by the Polyakov nonlocal action. Furthermore, two local terms with a
constant $\gamma$ are added in order to solve the model exactly. Then,
the usual semi-classical equations of motion obeying the conventional
Poisson algebra are solved, and the finite accelerating expansion is
obtained under some conditions; however, they do not exhibit any
change in phase. In sect.~\ref{sec:NC}, new equations of motion based
on the modified Poisson brackets are shown to give the nontrivial
solutions depending on the noncommutative parameter. For $\gamma >2$,
it is shown that the second accelerating expansion of the universe
appears smoothly from the decelerating phase which follows the initial
accelerating expansion. The second acceleration is finite and
eventually vanishes so that the infinite accelerating expansion or the
curvature singularity does not exist. Finally, some comments and
discussions follow in sect.~\ref{discuss}.

\section{permanent accelerating expansion}
\label{sec:C}
In this section, we study the two-dimensional semi-classical dilaton 
gravity action composed of the conformal matter fields and its quantum
correction, which are described by the nonlocal Polyakov action with
two covariant local terms as follows~\cite{cghs},
\begin{equation}
   S = S_\mathrm{DG} + S_\mathrm{cl} + S_\mathrm{qt}. \label{action:total}
\end{equation}
The first term on the right-hand-side is the well-known string
inspired dilaton gravity action written as
\begin{equation}
  S_\mathrm{DG} = \frac{1}{2\pi} \int d^2 x \sqrt{-g} e^{-2\phi} \left[
     R + 4 (\nabla \phi)^2 + 4 \lambda^2\right], \label{action:dg}
\end{equation}
and the classical matter action of $N$ conformal fields
$S_\mathrm{cl}$ and its one-parameter-family quantum correction
$S_\mathrm{qt}$ are given by
\begin{eqnarray}
  S_\mathrm{cl} &=& \frac{1}{2\pi} \int d^2 x \sqrt{-g} \left[-\frac12
     \sum_{i=1}^{N} (\nabla f_i)^2 \right], \label{action:cl} \\
  S_\mathrm{qt} &=& \frac{\kappa}{2\pi} \int \sqrt{-g} \left[ - \frac14
     R\frac{1}{\Box} R + (\gamma - 1) (\nabla\phi)^2 - \frac\gamma2
     \phi R \right], \label{action:qt}
\end{eqnarray}
and $\kappa = (N-24)\hbar/12$. The higher order quantum correction beyond the
one-loop is negligible in the large $N$ approximation where $N \to
\infty$ and $\hbar \to 0$, so that $\kappa$ is assumed to be positive
finite constant, while the cosmological constant $\lambda^2$ is set to
zero for simplicity. 
Note that the local ambiguity terms in Eq.~(\ref{action:qt}) correspond to
those of the Russo-Susskind-Thorlacius(RST) model~\cite{rst} for
$\gamma=1$, and to the Bose-Parker-Peleg model~\cite{bpp} for
$\gamma=2$. In our work, $\gamma$ is assumed to be $\gamma>2$ for our
goal.

In the conformal gauge, $ds^2 = -e^{2\rho} dx^+ dx^-$, if we define new
fields as follows
\begin{eqnarray}
\chi &=& e^{-2\phi} + \kappa \left(\rho - \frac\gamma2 \phi \right),  \label{new:chi} \\ 
\Omega &=& e^{-2\phi} - \frac\kappa2 (\gamma - 2) \phi,  \label{new:Omega}
\end{eqnarray}
the total action~(\ref{action:total}) is written as
\begin{equation}
S = \frac{1}{\pi} \int\/d^2 x \left[ -\frac1\kappa \partial_+ \chi
  \partial_- \chi + \frac1\kappa \partial_+ \Omega \partial_- \Omega
  + \frac12 \sum_{i=1}^N \partial_+ f_i \partial_- f_i \right], \label{action:new}
\end{equation}
and constraints are given by
\begin{equation}
\kappa t_\pm = - \frac1\kappa (\partial_\pm \chi)^2 + \partial_\pm^2
  \chi + \frac1\kappa (\partial_\pm \Omega)^2 + \frac12 \sum_{i=1}^N
  (\partial_\pm f_i)^2, \label{constraint:conf}
\end{equation}
where $t_\pm$ reflects the nonlocality of the anomaly term in the
Polyakov action. This integration function from the nonlocality
should be determined by the choice of the boundary condition for 
the geometrical vacuum and matter state, which may be constant or
time-dependent depending on model. The quantum energy-momentum tensors
from Eq.~(\ref{action:qt}) can be written in the form of
\begin{eqnarray}
  T^\mathrm{qt}_{+-} &=& \frac{\kappa(\gamma-2)}{2\Omega'} \left[
  \partial_+\partial_-\Omega - \frac{\Omega''}{(\Omega')^2}
  \partial_+\Omega \partial_-\Omega \right] -
  \partial_+\partial_-(\chi-\Omega), \label{energy:+-} \\
  T^\mathrm{qt}_{\pm\pm} &=& \frac{\kappa}{\Omega'} \left[
    \partial_\pm^2\Omega - \frac{\Omega''}{(\Omega')^2}
    (\partial_\pm\Omega)^2 \right] + \partial_\pm^2(\chi-\Omega) -
    \kappa \left[ \frac{\partial_\pm\Omega}{\Omega'} + \frac1\kappa
    \partial_\pm(\chi-\Omega) \right]^2 - \kappa t_\pm \nonumber \\
  & & - \frac{\kappa(\gamma-1)}{(\Omega')^2}(\partial_\pm\Omega)^2 -
    \frac{\kappa\gamma}{2\Omega'} \left[ \partial_\pm^2\Omega -
    \frac{\Omega''}{(\Omega')^2} (\partial_\pm\Omega)^2 \right],  \label{energy}
\end{eqnarray}
where $\Omega' = d\Omega/d\phi $ and $\Omega'' = 4 e^{-2\phi}$.
Assuming the homogeneous spacetime, the Lagrangian and the constraints
are reduced to
\begin{eqnarray}
  L &=& - \frac{1}{2\kappa} \dot{\chi}^2 + \frac{1}{2\kappa}
      \dot{\Omega}^2 + \frac14 \sum_{i=1}^N \dot{f}_i^2, \label{lagrangian} \\
  \kappa t_{\pm} &=& - \frac{1}{4\kappa} \dot{\chi}^2 + \frac14
      \ddot{\chi} + \frac{1}{4\kappa} \dot{\Omega}^2 + \frac18
      \sum_{i=1}^N \dot{f}_i^2,  \label{constraint}
\end{eqnarray}
where the Lagrangian is defined by $S/L_0 = \frac{1}{\pi} \int dt L$
with $L_0=\int dx$. The overdot denotes the
derivative with respect to the coordinate time $t$ defined by $dx^\pm
= dt \pm dx$.
Then, the Hamiltonian is 
\begin{equation}
  H = - \frac\kappa2 P_\chi^2 + \frac\kappa2 P_\Omega^2 + \sum_{i=1}^N P_{f_i}^2,
  \label{hamiltonian}
\end{equation}
where the canonical momenta are given by $P_\chi = - \frac1\kappa
\dot{\chi}$, $P_\Omega = \frac1\kappa \dot{\Omega}$, $P_{f_i} =
\frac12 \dot{f}_i$.

If we now define non-vanishing Poisson brackets between elementary
fields as follows
\begin{equation}
  \{\Omega, P_\Omega\}_{\mathrm{PB}} = \{\chi, P_\chi\}_{\mathrm{PB}} =
  \{f_i, P_{f_i}\}_\mathrm{PB} = 1,  \label{PB:C}
\end{equation}
then Hamiltonian equations of motion~\cite{bk:1}, $\dot{\cal O} = \{
{\cal O}, H \}_{\mathrm{PB}}$, for canonical fields are derived as
\begin{eqnarray}
  & & \dot\chi = - \kappa  P_\chi, \quad \dot\Omega = \kappa
    P_\Omega, \quad \dot{f}_i = 2 P_{f_i},  \label{eq:field:C} \\
  & & \dot{P}_\chi = \dot{P}_\Omega = \dot{P}_{f_i} = 0,  \label{eq:momentum:C}
\end{eqnarray}
which are solved as
\begin{eqnarray}
  & & \chi = -\kappa P_{\chi_0} t + \chi_0,  \label{sol:chi:C} \\
  & & \Omega = \kappa P_{\Omega_0} t + \Omega_0,  \label{sol:Omega:C}
\end{eqnarray}
where $P_{\chi_0}$, $P_{\Omega_0}$, $\chi_0$, and $\Omega_0$ are
arbitrary constants, and $P_{f_i}=0$ for the sake of simplicity. 
Note that these semiclassical solutions~(\ref{sol:chi:C}) and
(\ref{sol:Omega:C}) from the Hamiltonian equations of
motion~(\ref{eq:field:C}) and (\ref{eq:momentum:C}) are essentially
equivalent to those of Euler-Lagrangian equations of motion from
Eq.~(\ref{lagrangian}) because the fields $\Omega$ and $\chi$ are not
the quantum operators. If they are operators, they should be
decomposed into the positive and the negative frequency parts along
with the creation and annihilation operators. So, this is not the
quantization of the quantization for the model~(\ref{action:cl}).  
In the next section, we will modify the conventional Poisson brackets
in this semiclassical regime, in order to obtain the modified
semiclassical equations of motion.   

We now turn to the issue of the expanding universe. We first consider
the expansion condition for the scale factor,
\begin{equation}
 \frac{da}{d\tau} = \frac{d\rho}{dt} = \frac{\kappa
 P_{\Omega_0}}{-2e^{-2\phi} - \kappa(\gamma-2)/2} - (P_{\chi_0} +
 P_{\Omega_0}) \ge 0,  \label{velocity:C}
\end{equation}
where the scale factor $a(\tau)$ is a function of the comoving time
$\tau$ and is defined by $ds^2 = -d\tau^2 + a^2(\tau) dx^2$, that is,
$d\tau = e^{\rho(t)}dt$ and $a(\tau) = e^{\rho(t)}$. Since the
denominator in Eq.~(\ref{velocity:C}) is less than $-\kappa
(\gamma-2)/2$, we get the following condition,
\begin{equation}
P_{\chi_0} + \frac{\gamma}{\gamma-2} P_{\Omega_0} \le 0,
          \label{cond:expanding:C}
\end{equation}
where $P_{\Omega_0}>0$ is assumed. As for the constraints,
substituting the solutions~(\ref{sol:chi:C}) and (\ref{sol:Omega:C})
into the constraint equations~(\ref{constraint}) give 
\begin{equation}
  \kappa t_\pm = -\frac\kappa4 (P_{\chi_0}^2 - P_{\Omega_0}^2),
          \label{constraint:C}
\end{equation}
where $t_\pm$ is constant determined by the matter state. 
  
The curvature scalar is calculated as
\begin{equation}
  R =
    \frac{\kappa^2P_{\Omega_0}^2e^{-4\phi}}{[e^{-2\phi}+\kappa(\gamma-2)/4]^3}
    \exp\left[2(P_{\chi_0}+P_{\Omega_0})t-\frac2\kappa(\chi_0-\Omega_0)\right],
      \label{R:C}
\end{equation}
which can not be negative, since we have assumed that $\kappa$ is
positive and $\gamma>2$. Because the curvature scalar in two
dimensions is directly proportional to the second derivative of the
scale factor, Eq.~(\ref{R:C}) shows that the universe exhibits
permanent accelerating expansion without any decelerating
expansion. Note that the curvature scalar~(\ref{R:C}) converges to
zero at both ends if the following condition is met,
\begin{eqnarray}
  P_{\chi_0} + \frac{\gamma+2}{\gamma-2} P_{\Omega_0} > 0.  \label{cond:nonsingular:C}
\end{eqnarray}
Then, from the expansion condition~(\ref{cond:expanding:C}) and
convergence of the scalar curvature~(\ref{cond:nonsingular:C}), we
get 
\begin{equation}
-\frac{\gamma+2}{\gamma-2} P_{\Omega_0} < P_{\chi_0} 
\le - \frac{\gamma}{\gamma-2} P_{\Omega_0}.
\end{equation}
Thus, we have the singularity-free accelerating solution under this
condition. For other choices of the constants, what we get is the
decelerating solution which unfortunately does not show any phase
change of the universe. In the next section, by modifying the Poisson
brackets~(\ref{PB:C}), we will find another solution showing the
desired phase change to the accelerating expansion from decelerating
expansion following the initial exponentially accelerating expansion.

\section{phase changing universe} 
\label{sec:NC}
We now extend the conventional (commutative) Poisson brackets to the
modified (noncommutative) Poisson brackets characterized by two
noncommutative constants, $\Theta$ and $\theta$, which are reminiscent
of the one appearing in the noncommutative algebra of the D-brane on
the constant tensor field or a point particle moving very slowly on
the constant magnetic field~\cite{sw,vas,ko}.
In our case, we are trying to obtain the modified semiclassical
solution involving the noncommutative constants from the modified
semiclassical equations of motion based on the noncommutative algebra.
We now consider noncommutative case of the Poisson algebra as follows~\cite{bn}
\begin{eqnarray}
  & & \{ \Omega, P_\Omega \}_{\mathrm{MPB}} 
    = \{ \chi, P_\chi \}_{\mathrm{MPB}}
    = \{f_i, P_{f_i}\}_\mathrm{MPB} = 1, \nonumber \\ 
  & & \{ \chi, \Omega \}_{\mathrm{MPB}} = \Theta, \quad \{ P_\chi,
    P_\Omega \}_{\mathrm{MPB}} = \theta, \quad \mathrm{others} = 0, \label{PB:NC}
\end{eqnarray}
where $\Theta$ and $\theta$ are positive independent constants.
Then, modified semiclassical equations of motion are given by
\begin{eqnarray}
  & & \dot\chi = \{ \chi, H \}_{\mathrm{MPB}} = -\kappa P_\chi, \quad
      \dot\Omega =  \{ \Omega, H \}_{\mathrm{MPB}} = \kappa P_\Omega, 
      \label{eq:field:NC}\\
  & & \dot{P}_\chi =  \{ P_\chi, H \}_{\mathrm{MPB}} =\kappa \theta
      P_\Omega, \quad \dot{P}_\Omega =  \{ P_\Omega, H \}_{\mathrm{MPB}} =
      \kappa\theta P_\chi. \label{eq:momentum:NC}
\end{eqnarray}
Note that the original semiclassical equations of
motion~(\ref{eq:field:C}) and (\ref{eq:momentum:C}) are recovered in
the limit, $\theta \rightarrow 0$. These modified semiclassical
equations of motion depend only on $\theta$, since the
Hamiltonian~(\ref{hamiltonian}) is described by the pure momentum
variables. Combining Eqs.~(\ref{eq:field:NC}) and
(\ref{eq:momentum:NC}), we get
\begin{equation}
  \ddot{P}_\chi = \kappa^2 \theta^2 P_\chi, \quad \ddot{P}_\Omega =
  \kappa^2 \theta^2 P_\Omega, \label{eq:reduced:NC}
\end{equation}
and their solutions are obtained as
\begin{eqnarray}
  \chi &=& -\alpha \sinh \kappa\theta t
     - \beta \cosh \kappa\theta t + C_\chi,  \label{sol:chi:NC} \\
  \Omega &=& \alpha \cosh \kappa\theta t 
     + \beta \sinh \kappa\theta t + C_\Omega,  \label{sol:Omega:NC}
\end{eqnarray}
where $\alpha$, $\beta$ and $ C_\chi$, $C_\Omega$ are integration 
constants.

\begin{figure}
\includegraphics{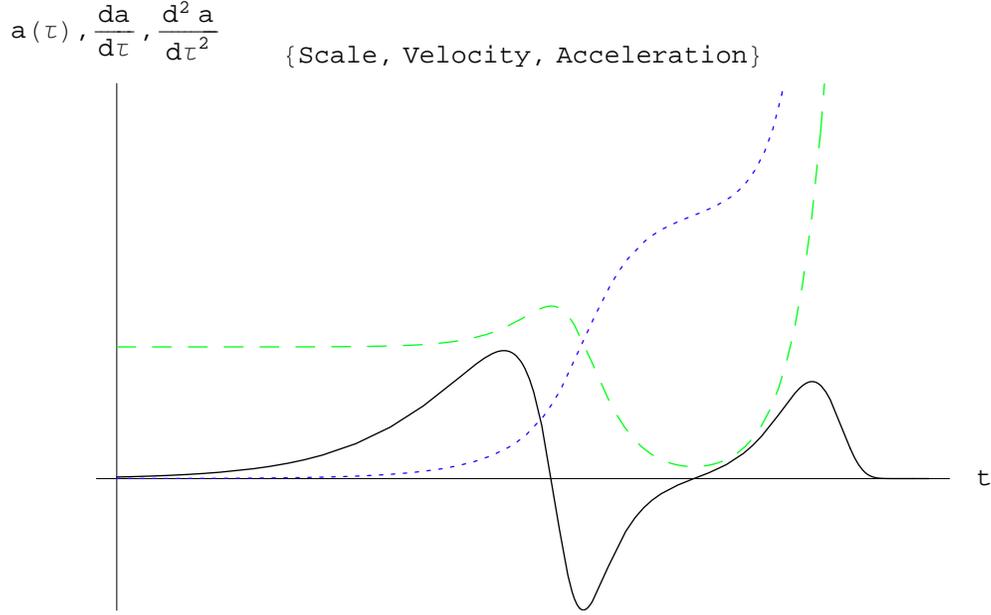}
\caption{\label{fig:scale} 
The dotted and dashed lines show the behavior of the scale factor $a(\tau)$
and the expansion rate $da/d\tau$, respectively.
The solid line importantly means the profile for the acceleration,
$d^2a/d\tau^2$. The area of the scale factor in the figure
represents the comoving time $\tau$, so that the scale will not blow
up in a finite comoving time $\tau$. 
It is shown that the first acceleration starts from the comoving time
$\tau=0$ corresponding to the past infinity of $t \rightarrow
-\infty$, and then the deceleration of the universe corresponding to
FRW phase appears in a finite time. Subsequently, the
second acceleration turns up. In this figure, the parameters and
constants are set as $\kappa=1$, $\gamma=18$,
$\theta=1$, $\alpha=0$, $\beta=-0.1$, $C_\chi=C_\Omega=1$, and
$P_{f_i}=0$.}
\end{figure}

From the solutions~(\ref{sol:chi:NC}) and (\ref{sol:Omega:NC}), we can
obtain the expansion condition at the asymptotic region, which comes from
\begin{equation}
  \frac{da}{d\tau} = \frac{-\kappa\theta/2}{e^{-2\phi} +
  \kappa(\gamma-2)/4} (\alpha \sinh \kappa\theta t + \beta \cosh
  \kappa\theta t) - \theta (\alpha+\beta) e^{\kappa\theta t} > 0.  \label{velocity}
\end{equation}
At the asymptotic future infinity and the past infinity,
$t\to\pm\infty$, the following inequalities can be derived,
\begin{equation}
  \alpha + \beta < 0, \quad \alpha - \beta > 0,  \label{cond:expanding:NC}
\end{equation}
where these conditions imply that the constant $\beta$ is negative.
In the intermediate region it is not easy to write down the condition
in a simplified form; however, it can be shown in Fig.~\ref{fig:scale}
that the positive expansion rate depicted in terms of the dashed line
is possible without contraction of the universe.

Let us remind that some of dark-energy-dominant accelerating models
have a defect of so-called big rip singularity that the scale blows up in a
finite time~\cite{ckw} (for a recent review, see
Ref.~\cite{cst}). Since the present model is different from the
previous models, we investigate whether this kind of singularity
appears or not. If we rewrite the scale factor as
\begin{equation}
  a(\tau) = e^\rho = e^\phi \exp\left[ -\frac1\kappa (\alpha +
    \beta) e^{\kappa\theta t} + \frac1\kappa (C_\chi - C_\Omega)
  \right],  \label{scale}
\end{equation}
then we see that it is definitely finite except it becomes
infinite, $e^{\rho} \sim \exp[-\frac{\gamma-1}{\kappa(\gamma-2)}
(\alpha+\beta) e^{\kappa\theta t}]$, only for the coordinate time
$t\to\infty$ by using Eqs.~(\ref{new:Omega}), (\ref{sol:Omega:NC}),
and (\ref{cond:expanding:NC}). The infinite coordinate time is
related to the infinite comoving time as $\tau(t \rightarrow \infty) \sim
\infty$ and $\tau(t \rightarrow -\infty) \sim 0$, and the scale factor
is finite at a finite comoving time. 
Note that as $t \to - \infty$, the dilaton field is approximated as 
$e^{-2\phi} \sim e^{-\kappa \theta t}$ from Eqs.~(\ref{new:Omega})
and (\ref{sol:Omega:NC}), and then the scale is given 
by $a(\tau) \sim e^{\kappa \theta t/2}$, which is reminiscent of the 
initial inflation.

Next, we investigate the behavior of the curvature scalar,
\begin{eqnarray}
  R &=& \frac{2d^2a/d\tau^2}{a(\tau)} \nonumber \\
  &=& e^{-2\phi} \exp \left[ \frac2\kappa (\alpha +
    \beta) e^{\kappa\theta t} - \frac2\kappa (C_\chi - C_\Omega)
  \right] \times \nonumber \\
  & & \times \bigg\{ -\frac{\kappa^2\theta^2}{e^{-2\phi} +
    \kappa(\gamma-2)/4} \bigg[ (\alpha \cosh \kappa\theta t + \beta
    \sinh \kappa\theta t) - \nonumber \\
  & & \qquad - \frac{e^{-2\phi}}{[e^{-2\phi} +
    \kappa(\gamma-2)/4]^2} (\alpha \sinh \kappa\theta t + \beta \cosh
    \kappa\theta t)^2 \bigg] - 2\kappa\theta^2 (\alpha + \beta)
    e^{\kappa\theta t} \bigg\}.  \label{R:NC}
\end{eqnarray}
Note that $R \simeq -\kappa^4\theta^3(\gamma-2)t/4 \to +\infty$ for the
limit of $t\to-\infty$, and $R \sim e^{\kappa\theta t} \exp
[\frac{2(\gamma-1)}{\kappa(\gamma-2)} (\alpha+\beta) e^{\kappa\theta
  t}] \to 0$ for $t\to\infty$,
where we have employed $\alpha + \beta < 0$ in
Eq.~(\ref{cond:expanding:NC}). This model is singularity-free
everywhere except the initial singularity at $\tau=0$($t\to-\infty$).
Therefore, there is no (big rip) singularity in a finite future comoving time.

Let us now study the most intriguing issue of the
late acceleration. The acceleration is calculated formally as
\begin{eqnarray}
  \frac{d^2a}{d\tau^2} &=& \ddot{\rho} e^{-\rho} \nonumber \\
  &=& \frac12 e^{-\phi} \exp \left[ \frac1\kappa (\alpha +
    \beta) e^{\kappa\theta t} - \frac1\kappa (C_\chi - C_\Omega)
  \right] \times \nonumber \\
  & & \times \bigg\{ -\frac{\kappa^2\theta^2}{e^{-2\phi} +
    \kappa(\gamma-2)/4} \bigg[ (\alpha \cosh \kappa\theta t + \beta
    \sinh \kappa\theta t) - \nonumber \\
  & & \quad - \frac{e^{-2\phi}}{[e^{-2\phi} +
    \kappa(\gamma-2)/4]^2} (\alpha \sinh \kappa\theta t + \beta \cosh
    \kappa\theta t)^2 \bigg] - 2\kappa\theta^2 (\alpha + \beta)
    e^{\kappa\theta t} \bigg\}.  \label{acceleration}
\end{eqnarray}
Note that it vanishes at both ends, $t\to\pm\infty$, which is asymptotically
given in the form of $d^2a/d\tau^2 \sim -te^{\kappa\theta t} $ for
$t\to-\infty$ and $d^2a/d\tau^2 \sim
\exp[\frac{\gamma-1}{\kappa(\gamma-2)} (\alpha+\beta)
e^{\kappa\theta t}]$ for $t\to\infty$.   
According to the assumption $\gamma >2$ and $\alpha +\beta < 0$ in
Eq.~(\ref{cond:expanding:NC}), there does not exist any
divergent acceleration at both ends even though the initial curvature
singularity appears. Apart from the ends, in the intermediate region,
the whole profile of the acceleration~(\ref{acceleration})
is plotted in Fig.~\ref{fig:scale}.
It shows that the universe starts with the inflation era and changes its state to
Friedmann-Robertson-Walker(FRW) phase, and then it remarkably ends up
with the desired second acceleration.

In order to discuss the equation of state parameter, we first set the
energy-momentum tensors as a source by using the constraint
equations~(\ref{constraint}),
\begin{eqnarray}
  T_{\pm\pm}&=&-\kappa t_\pm \\ 
    &=& \frac{1}{4\kappa} \left(\alpha^2 - \beta^2\right) +
  \frac{\kappa^2\theta^2}{4} \left(\alpha\sinh\kappa\theta t +
  \beta\cosh\kappa\theta t\right).
  \label{constarint:NC}
\end{eqnarray}
Then the energy density $\varepsilon$ and the pressure $p$ in the
comoving coordinates are
\begin{eqnarray}
\varepsilon &=& T_{\tau\tau} = e^{-2\rho} \left[
  T_{++} + 2T_{+-} + T_{--}
  \right], \label{density}  \\
p &=& T_{xx} = \left[ T_{++} -
  2T_{+-} + T \right].  \label{pressure}
\end{eqnarray}
Because of $T_{+-} = 0$, the equation-of-state parameter is 
simply $e^{2\rho}$, which is explicitly given by
\begin{eqnarray}
  \omega &=& p/\varepsilon \nonumber \\
  &=& e^{2\phi} \exp \left[ -\frac2\kappa (\alpha +
    \beta) e^{\kappa\theta t} + \frac2\kappa (C_\chi - C_\Omega)
  \right].
\end{eqnarray}
\begin{figure}
\includegraphics{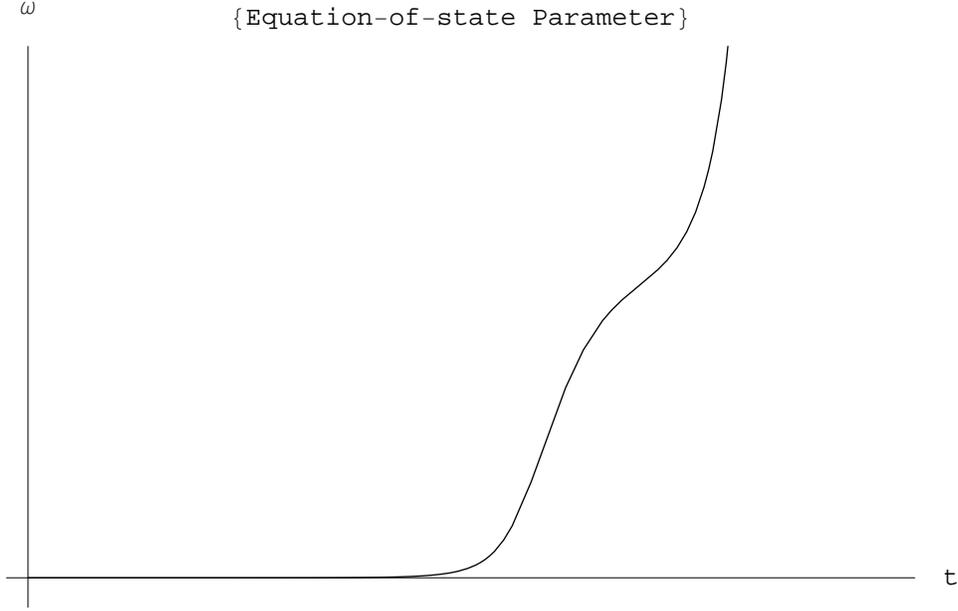}
\caption{\label{fig:omega}
The solid line shows the behavior of the equation-of-state parameter
$\omega$, where the various parameters in Fig.~\ref{fig:scale} are used.} 
\end{figure}

It is monotonically increasing and then eventually diverges because
of the vanishing energy density. Note that both the energy density and
the pressure are always negative so that the state parameter is
positive, which is different from the previous models. In the past,
the state parameter is finite because the energy
density and the pressure are all negative infinity. As time goes on,
the energy density becomes finite while the pressure goes to the
negative infinity. The whole profile for the state parameter is shown
in Fig.~\ref{fig:omega} along with the same fixed parameter used in
Fig.~\ref{fig:scale}.

The energy-momentum tensor taken here is a little bit different from
the conventional treatment of the energy-momentum tensors in that the
geometrical part and the source part are not distinguished, since our
energy-momentum tensors are induced from the vacuum polarization which
is expressed by the metric and the dilaton without the classical
matter. The induced source terms are combined with the original
metric part and hence result in the back-reacted metric. The remnant,
that is $t_\pm$, is interpreted as a source of this model, 
which is seen in Eq.~(\ref{constraint:conf}). The left is a
source while the right-hand-side is the back-reacted metric part.

\section{Discussion}
\label{discuss}
In our model, the induced stress-tensors from the one-loop effective
action~(\ref{action:qt}) play an important role. 
It consists of the integration function $t_\pm$ and the bulk part in
Eqs.~(\ref{energy}), where the former function is interpreted as a source
contribution, while the latter part was solved in cooperation with the
metric and dilaton coupled manner for the quantum back reaction 
as seen in Eq.~(\ref{constraint:conf}). 
This means that the scale factor from $\Omega$ and $\chi$ 
reflects the back reaction of the geometry.
So, the classical dilaton-gravity part and the quantum-mechanically
induced part have not
been separated; instead, they have been solved in terms of the
redefined fields $\Omega$ and $\chi$. 
     
The modified Poisson brackets~(\ref{PB:NC}) look
like noncommutative commutator if we define the relevant fields
as operators~\cite{sw}. However, 
in our case, we have just defined Poisson brackets in order to obtain 
modified semiclassical equations of motion which recover the original
semiclassical equations of motion when the noncommutative parameter
$\theta$ vanishes. In the quantum mechanics, this nontrivial Poisson
brackets may appear in a slowly moving electron on the constant
magnetic field~\cite{ko}; unfortunately, we have no idea what the
counterpart is in this cosmological model.

In fact, we have addressed the same issue in the RST model 
in Ref.~\cite{ky:ncdc}. In that case, it shows just a single
phase transition from the decelerating to the accelerating universe
and, even worse, the future singularity appears. However, 
in this improved model, we 
obtained the future-singularity-free second accelerating expansion
along with some local ambiguity terms for the Polyakov non-local action.
And, we skipped the critical case of $\gamma=2$ which is similar
to our model of $\gamma >2$.


\begin{acknowledgments}
We would like to thank R. H. Brandenberger for helpful comments and
M. S. Yoon for discussions.
W.~Kim was supported by the Science Research Center Program of the
  Korea Science and Engineering Foundation through the Center for
  Quantum Spacetime \textbf{(CQUeST)} of Sogang University with grant
  number R11 - 2005 - 021.
E.~J.~Son was supported by the Korea Research Foundation Grant funded
by Korea Government(MOEHRD, Basic Research Promotion Fund)
(KRF-2005-070-C00030).
\end{acknowledgments}



\begin{thebibliography}{99}
\bibitem{perlmutter} S.~Perlmuttter {\it et al.},
  Astrophys.\ J.\ \textbf{517}, 565(1999) [astro-ph/9812133];
M.~S.~Turner, \textit{Why cosmologists believe the
  universe is accelerating}, astro-ph/9904049;
  D.~N.~Spergel {\it et al.},
  Astrophys.\ J.\ Suppl.\ \textbf{148}, 175 (2003)
  [astro-ph/0302209].

\bibitem{caldwell} R.~R.~Caldwell, Phys.\ Lett.\ B \textbf{545}, 23
  (2002) [astro-ph/9908168].

\bibitem{carroll} S.~M.~Carroll, \textit{Why is the universe
    accelerating?}, [astro-ph/0310342];
P.~S.~Apostolopoulos and N.~Tetradis, Phys.\ Lett.\ B
  \textbf{633}, 409 (2006) [hep-th/0509182];
T.~Kobayashi, Phys.\ Rev.\ D \textbf{73}, 124031
  (2006) [hep-th/0602168];
V.~Faraoni, M.~N.~Jensen, and S.~A.~Theuerkauf,
  Class.\ Quant.\ Grav.\ \textbf{23}, 4215 (2006) [gr-qc/0605050];
I.~Dymnikova and M.~Fil'chenkov, Phys.\ Lett.\ B
  \textbf{635}, 181 (2006) [gr-qc/0606048];
H.~Wei and R.-G.~Cai,
  Phys.\ Rev.\ D \textbf{73}, 083002 (2006) [astro-ph/0603052].

\bibitem{cghs} C.~G.~Callan, S.~B.~Giddings, J.~A.~Harvey, and
  A.~Strominger, Phys.\ Rev.\ D {\bf 45}, R1005 (1992)
  [hep-th/9111056].

\bibitem{str}
  A.~Strominger,
  Phys.\ Rev.\ D {\bf 46}, 4396 (1992)
  [arXiv:hep-th/9205028].

\bibitem{rst} J.~G.~Russo, L.~Susskind, and L.~Thorlacius,
  Phys.\ Rev.\ D {\bf 46}, 3444 (1992) [hep-th/9206070].

\bibitem{mrcm} R.~B.~Mann and S.~F.~Ross, Phys.\ Rev.\ D \textbf{47},
  3312 (1993) [hep-th/9206022]; K.~C.~K.~Chan and R.~B.~Mann,
  Class.\ Quant.\ Grav.\ \textbf{10}, 913 (1993) [gr-qc/9210015].

\bibitem{bpp} S.~Bose, L.~Parker, and Y.~Peleg,
  Phys.\ Rev.\ Lett.\ \textbf{76}, 861 (1996) [gr-qc/9508027].

\bibitem{rt} J.~Russo and A.~Tseytlin, Nucl.\ Phys.\ B \textbf{382}, 259 (1992).

\bibitem{kv} W.~Kummer and D.~V.~Vassilevich, Phys.\ Rev.\ D \textbf{60},
  084021 (1999) [hep-th/9811092].

\bibitem{bc} A.~Bilal and C.~Callan, Nucl.\ Phys.\ B \textbf{394}, 73
  (1993) [hep-th/9205089]. 

\bibitem{bk:1} A.~Bilal and I.~I.~Kogan, Phys.\ Rev.\ D \textbf{47}, 5408
  (1993) [hep-th/9301119]. 

\bibitem{gv}  M.~Gasperini and G.~Veneziano, Phys.\ Lett.\ B {\bf 387},
  715 (1996) [hep-th/9607126]. 

\bibitem{rey}   S.~J.~Rey, Phys.\ Rev.\ Lett.\ {\bf 77}, 1929 (1996)
  [hep-th/9605176].

\bibitem{bk} S.~Bose and S.~Kar, Phys.\ Rev.\ D \textbf{56}, R4444
  (1997) [hep-th/9705061].

\bibitem{ky:dg} W.~Kim and M.~S.~Yoon, Phys.\ Lett.\ B \textbf{423},
  231 (1998) [hep-th/9706154]. 

\bibitem{ky:bd} W.~Kim and M.~S.~Yoon, Phys.\ Rev.\ D \textbf{58},
  084014 (1998) [hep-th/9803081]. 

\bibitem{cdkz} M.~H.~Christmann, F.~P.~Devecchi, G.~M.~Kremer, and
  C.~M.~Zanetti,  Europhys.\ Lett.\ \textbf{67}, 728, (2004)
  [gr-qc/0407029].
 
\bibitem{no}
  S.~Nojiri and S.~D.~Odintsov,
  Int.\ J.\ Mod.\ Phys.\ A {\bf 16}, 1015 (2001)
  [arXiv:hep-th/0009202].

\bibitem{ky:ncdc} W.~Kim and M.~S.~Yoon, \textit{Accelerating universe in
  two-dimensional noncommutative dilaton cosmology}, gr-qc/0608032.

\bibitem{sw} N.~Seiberg and E.~Witten, J.\ High Energy
  Phys.\ \textbf{09}, 032 (1999) [hep-th/9908142].

\bibitem{vas} D.~V.~Vassilevich, \textit{Stability of a noncommutative
    Jackiw-Teitelboim gravity}, hep-th/0602095 (2006). 

\bibitem{ko} W.~Kim and J.~J.~Oh, Mod.\ Phys.\ Lett.\ A \textbf{15}, 1597
  (2000) [hep-th/9911085].

\bibitem{bn} G.~D.~Barbosa and N.~Pinto-Neto, Phys.\ Rev.\ D
  \textbf{70}, 103512 (2004) [hep-th/0407111].

\bibitem{ckw}
  R.~R.~Caldwell, M.~Kamionkowski, and N.~N.~Weinberg,
  Phys.\ Rev.\ Lett.\  {\bf 91}, 071301 (2003)
  [astro-ph/0302506];
  S.~M.~Carroll, M.~Hoffman, and M.~Trodden,
  Phys.\ Rev.\ D {\bf 68}, 023509 (2003)
  [astro-ph/0301273];
  S.~Nojiri and S.~D.~Odintsov,
  Phys.\ Lett.\ B {\bf 595}, 1 (2004)
  [hep-th/0405078];
  S.~Nojiri and S.~D.~Odintsov,
  Phys.\ Rev.\ D {\bf 70}, 103522 (2004)
  [hep-th/0408170];
  S.~Nojiri, S.~D.~Odintsov, and S.~Tsujikawa,
  Phys.\ Rev.\ D {\bf 71}, 063004 (2005)
  [hep-th/0501025].

\bibitem{cst}
  E.~J.~Copeland, M.~Sami, and S.~Tsujikawa,
  \textit{Dynamics of dark energy},
  hep-th/0603057.

\end{thebibliography}
\end{document}